\documentclass[12pt]{article}

\usepackage{graphicx,subfigure}
\usepackage{subfigure}
\usepackage{algorithm}
\usepackage{algorithmicx}
\usepackage{algpseudocode}
\usepackage{url}
\usepackage{lipsum}

\begin{document}
\title{Some Basic Radio System OPSEC Considerations}
\author{Joshua Davis\\ jmd@covert.codes}
\maketitle

\section{Introduction}
Radio Frequency (RF) channels and networks are often used to carry information whose existence may be necessary for organizational operation, or whose content must remain inaccessible to third parties.  Cryptographic solutions are commonly employed to render the information on RF links unintelligible to unauthorized recipients, though the existence and nature of the links themselves may assist an adversary in conducting attacks on the system, such as behavioral analysis or denial of service.  So, it is prudent that radio system administrators implement RF security plans that are broad in scope, not being limited to only cryptographic solutions, but also including means of foiling other forms of analysis (e.g. open-source, visual).  As in other arenas, Operations Security (OPSEC) is integral to the security of the system as a whole.

Here we will consider a few ways in which an adversary may gather information about our RF systems and resources.  Some techniques are universally applicable, perhaps with minor modifications, and some are applicable only to certain types of systems.  The information here should be used in conjunction with other security practices to approach {\em total system security}.

This is an unscientific introduction to basic radio frequency system OPSEC aspects that I have found to be overlooked and lacking in high security system deployments that may have benefited from them.  Please email me at the address above with any comments, corrections, or suggested additions.

\section{Cryptographic Security}
A common and inappropriate way to protect information on a wireless network is {\em obfuscation}.  Such ``protection'' relies on the assumption that some attribute of the wireless communication protocol makes access by unintended recipients prohibitively difficult.  Obfuscation of the signal may be intentional, or it may be a side effect of the technology used for the system.  Police departments have increasingly moved toward digital trunking systems for their radio communication.  These systems were not extremely accessible to the general public, so listening in on the networks required time and effort. Such ``security'' fails for two reasons.  First, a determined and knowledgeable adversary will have little trouble overcoming the difficulties imposed by the protocol, and accessing the communication.  Second, technology rapidly advances, and such protocol based hurdles are continually lowered.  To reference our previous example, commercial digital trunking radio receivers (both hardware and software) have become increasingly available and affordable, and are capable of decoding many common protocols.

Encryption, properly implemented, is effective in obscuring the content of communications.  Where encryption fails, implementation is generally to blame, as was the case in the very well known Wired Equivalent Privacy (WEP) security weaknesses \cite{Fluhrer01}.

Encryption occurs on many levels, with the level of cryptographic security generally increasing with the number of layers.  A hurdle to using encryption at each communication endpoint (e.g. computer) and wireless node is key distribution.  In high security environments, encryption is generally done at the border of each internal network (for example, at the network edge routers), as well as at the RF system endpoints.  The cryptographic keys may differ for each system and link, and should be changed regularly.  In the case of a spoke-hub system topology as is commonly found in satellite communication (SATCOM) networks, each node uses a unique key to communicate with the gateway (hub).

In United States Department of Defense lexic, the encrypted portion of a network is referred to as {\em black}, and the unencrypted portion as {\em red}.  Black and red system components and links are physically isolated so that electromagnetic leakage will not compromise the unencrypted communication.  Often, the red equipment is placed in its own secure facility or room.

Many systems implement a {\em session key}.  A session key is used once, and securely discarded once after the session comes to an end.  Such practice ensures that should an adversary recover one key, they will only be able to decode a relatively small amount of their intercepted encrypted communications.

Cryptography is very useful for protecting the contents of radio communications, if properly implemented, if a sufficiently strong cryptographic algorithm is used, and if good key management practices are adhered to.  But as we have mentioned, sometimes the nature of the wireless communications assist an adversary in implementing an attack against a system.  So, one will want to minimize the amount of information that is easily available about their systems, even when good encryption is used.
  
\section{Open-source Analysis}
Open-source intelligence may be provided by governing entities for radio systems that meet certain criteria (e.g. that exceed a certain height, or belong to a certain class of operation).  In the United States the Federal Communications Commission (FCC) Universal Licensing System (ULS) \cite{ULS} and Antenna Structure Registration (ASR) \cite{ASR} databases provide information on many systems and antennas.  This information includes physical location, registrant details, frequencies, and other information which may be useful in enumerating the systems controlled by a target and their locations, or identifying the owner of a given resource (e.g. antenna).

If one must register their radio systems with a public database, it is advisable to use administrative, rather than facility or antenna, addresses where possible.  The amount of information included in public databases should be constrained to the legal minimum, where OPSEC is a prime motivator.  An inquiry into the FCC ASR database is shown in Figure 1.

\begin{figure*}[ht]
  \centering
  \includegraphics[width=1.00\textwidth]{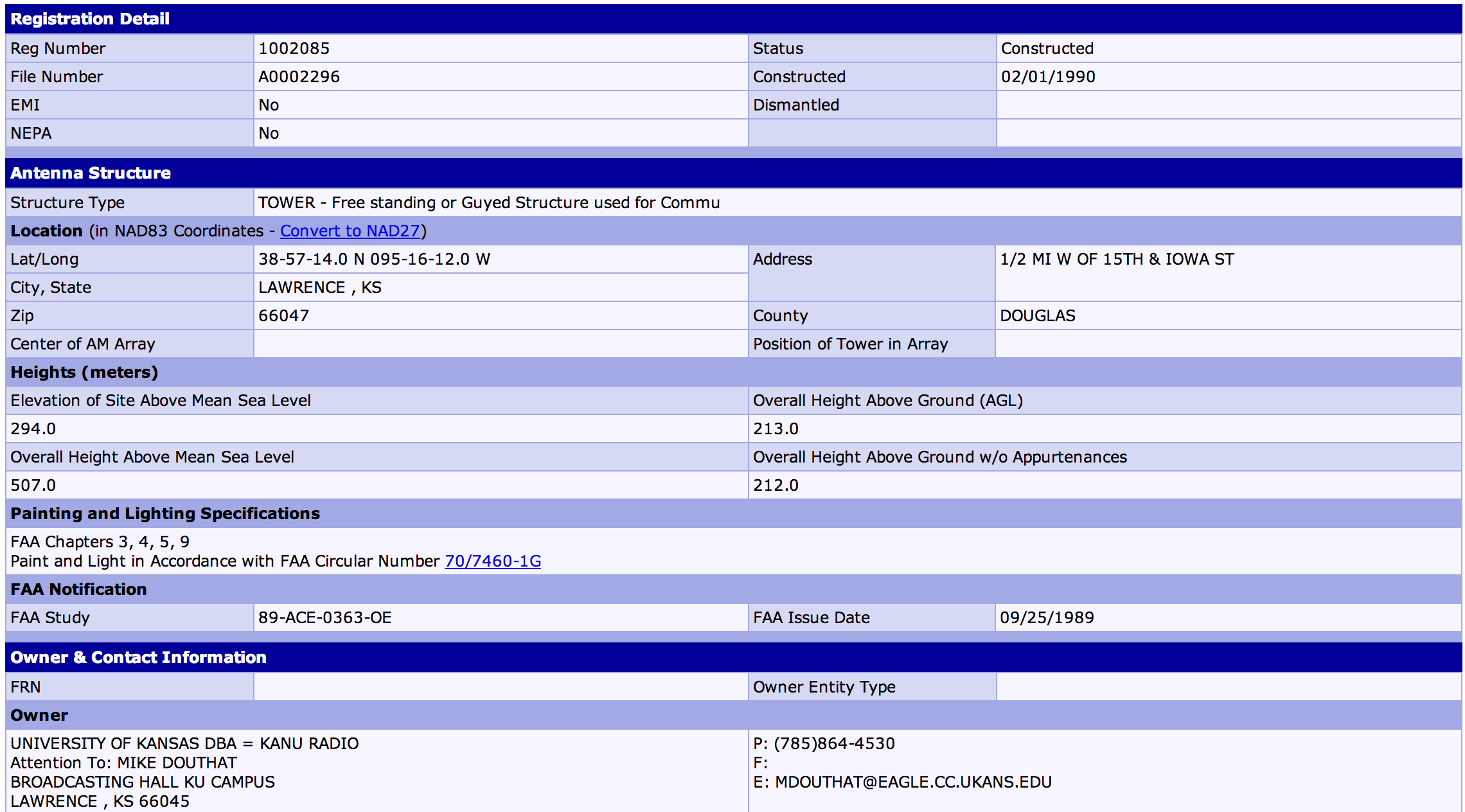}
  \caption{An ASR Database Record}
\end{figure*}

\section{Imagery and Visual Analysis}
Using tools such as Google Earth or some based on the Geographic information system (GIS), one may find and to some degree identify radio system components.  Buildings housing high-power radio equipment require the support of large air handling units, and often have large generators next to the main structure.  Such clues can help in identifying which facilities are critical to the radio system under analysis, once an adversary has narrowed down the physical area in which the system resides.  Antennas, including parabolic dishes, are often visible, and their size (circumference, length, or height) may be estimated by using online tools, or overhead land maps provided by local government (e.g. county).

An adversary can infer many things about a radio system by simply seeing the antenna attached to it.  The type, size, and orientation of an antenna reveal a lot about the system, as do logos and other markings.

Parabolic antennas, also known as {\em dishes} or {\em satellite dishes}, have two uses in communications: line-of-site, and satellite communication systems.  Those used for line-of-site communication point directly to another parabolic antenna, which may be several kilometers away.  The range at which line-of-site communication may occur is limited by the radio horizon, which is approximately $3.57 \cdot \sqrt{h}$ kilometers, where $h$ is the height of the antenna in meters; however, the range is often further limited by other propagation factors such as diffraction, reflection, and atmospheric attenuation.  In general, the direction in which an antenna points, along with knowledge of its radio horizon, may allow us to determine a direction and approximate maximum range for the antenna.

Dishes that do not point approximately horizontal may be part of a satellite communication system.  The gain of a parabolic antenna is proportional to its size; the size does not directly identify the frequency of the system.  Regardless, we can infer some facts about the system by the size of the dish.  Larger dishes are used where more gain is required, so systems with big parabolic antennas might be communicating with geosynchronous satellites, and may be transmitting as well as receiving.

When determining if a dish is part of a line-of-site or satellite communications system, one should take care to realize that the inclination of a satellite dish depends on the location of the satellite and on the latitude of the dish.  Further, many dishes use an offset configuration, where the feed horn is not located directly above the center of the dish, and the physical inclination does not correspond to the RF inclination.  So, it is possible to come across satellite antennas that appear to be pointed quite horizontally.  Dishes that do not move, or move very little, are communicating with geosynchronous satellites or are a line-of-site endpoint.  To determine if a stationary dish that seems to be pointing horizontal is a satellite or a line-of-site antenna, first calculate the geosynchronous look angle for your latitude, and reference the caveats in the previous paragraphs.

The frequencies used in satellite and line-of-site communication systems vary, but are often between the Ultra high frequency (UHF) and Extremely high frequency (EHF) bands.  If a waveguide is visible, its size can assist in narrowing down the frequency range of an antenna: bigger waveguides correspond to lower frequencies.  However, visual identification of waveguides in terms of frequency takes some practice.

We will now consider antennas other than parabolic dishes.  Antennas vary greatly in size, configuration, and function.  Understanding the function of an antenna, which may be derived from its physical attributes, may provide information about the system connected to it, its intended recipient(s), and perhaps its purpose.  In many antennas the size of the antenna, or at least its driven element, is a fraction (often $1/4$), of the operating wavelength of the system attached to it.  Each antenna has an associated radiation pattern and bandwidth, which may be approximated by referencing data sheets for antennas of the same type.  Some antennas are very directional, and others approach omnidirectional transmission.  Visual examination of an antenna provides information as to its directionality and orientation.  For antennas that operate locally (e.g. within the confines of an urban area), as may be determined by the operating frequency of the system (clues given by antenna size), high directionality may imply a point-to-point system configuration.  For antennas that operate over a large geographic area by taking advantage of High Frequency (HF) shortwave propagation, high directionality at the transmitter generally means a broad coverage area far from it.  Antennas that approximate omnidirectional propagation exist for the cases where the transmitter(s) or receiver(s) are mobile, or where there are multiple geographically diverse transmitter(s) or receiver(s).

Directionality affects the susceptibility of a radio system's communication to interception.  As previously mentioned, interception of a signal may be useful not only due to the information the signal contains, but also for behavioral analysis (if countermeasures are not implemented to counteract it), identification of the target's systems and  facility locations, and possibly, denial of service.  Interception may be easy, as is the case when the signal to be observed is being sent from a satellite (the signal is available throughout the satellite's footprint), or difficult (the target system utilizes highly directional antennas.)  If interception is difficult, one might still be able to intercept transmitter's communications by utilizing techniques given in the next section.

If overhead imagery is available, one may be able to approximate the size of dishes and other antennas.  If one has visual access to the transmitting antenna, they may be able to discern unique visual identifiers such as logos, that can be used in conjunction with the physical identifiers already mentioned (size, construction) to identify the antenna.  Once identified, information such as frequency, bandwidth, and side lobe power are readily available online.

It is good OPSEC to protect  antennas from visual observation, when the application or recipient of the associated system is secret.  One example of how this may be done is by covering the antenna with a radiation dome (Radome).  Note that one must protect their system from observers on the ground, as well as from overhead view by systems such as Google Earth.

\begin{figure}[h!]
  \centering

   \subfigure[A parabolic dish \cite{dish}]{
    \includegraphics[scale=1]{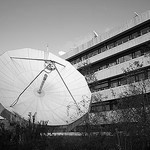}
  }
    \subfigure[A log-periodic directional antenna \cite{lpda}]{
    \includegraphics[scale=1]{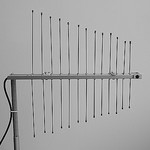}
  }\\
    \subfigure[Directional panel antennas, as used for Wi-Fi or cellular telephone \cite{panel}]{
    \includegraphics[scale=4.36]{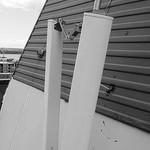}
  }
    \subfigure[Large omnidirectional broadcast antenna \cite{omni}]{
    \includegraphics[scale=1]{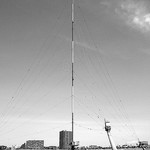}
  }
  
  \caption{Some common antenna types}
\end{figure}

\section{Advanced Interception}
Without direct access to the main transmission beam, one must rely on other methods to obtain access to the signal.  Electromagnetic waves have characteristics that differ by frequency.  For example, SATCOM relies on frequencies in bands that pass with acceptable attenuation through the atmosphere, while shortwave radio takes advantage of the properties of lower frequency (HF) radio waves to allow for geographically distant communication without satellites.

Some properties that may allow an adversary to indirectly access an RF signal are properties that are familiar from our understanding of light: reflection and diffraction.  Quantifying these effects in a given environment, with a unique set operating parameters and obstacles (which cause the reflection and diffraction) is likely prohibitively difficult, and in many cases such effects will be very minimal or perhaps even non-existent, where an antenna has a high look angle or no obstructions exist in the radiation path.  Having determined the frequency at which the antenna operates, an adversary may simply wander around the area around the transmitter ({\em around} being relative to transmit power and frequency), to determine if usable stray RF energy from the target system is accessible.  Keep in mind that it may be possible for an attacker to create an obstruction to reflect some transmit energy to their own equipment, if the transmit beam is reachable.

Antennas transmit in a primary direction, referred to as the {\em main lobe}.  Natural, though generally undesirable lobes of energy exist outside of the main lobe.  These occur at predictable intervals, and are referred to as {\em side lobes}.  Reference the Internet for information on a particular antenna's side lobe configuration.  It may be possible to place a receiver in a side lobe to receive some transmitted energy.  The strength of the side lobes are dependent on the antenna and transmit power.

To increase the difficulty an adversary will have intercepting an RF signal through reflection, diffraction, or by placing a receiver in a side lobe, one should plan their network so that antennas are placed as high as is practical, are pointed in a direction free from obstructions, and are as directive as is practical for the application.  Finally, transmit power should be kept as low as possible, while maintaining acceptable link characteristics such as signal-to-noise ratio).

\section{Conclusion: Pulling it All Together}
We have focused primarily on an adversary obtaining access to our RF energy, most of the time ignoring whether or not our communication is encrypted.  In the case of unencrypted communication, or encrypted communication with unencrypted control (protocol) information, the utility of interception depends on the value the adversary places on the communication or control information.  For links that are entirely encrypted, access to the RF energy may provide the attacker with information to assist in behavioral or geographical analysis of our systems.  For example, if transmitters are on when communication is taking place, and off when communication is not, an attacker can use access to the encrypted channel to help determine when high levels of user activity are taking place.  The directional orientation and characteristics (e.g. frequency, power) of antennas may allow the adversary to discern approximate location of nodes, as well as what sort of systems are being used.  The type of system being used may assist in determining the purpose of the communication supported by the system.

The level of paranoia (i.e. OPSEC) necessary for a given system or network is dependent on the value placed on the system and the information it contains.  Value of an individual system decreases as the number and quality of backup systems increases.  Security controls are subjective and one must weigh the costs and benefits when making a determination concerning which security controls to implement.

\bibliographystyle{plain}
\bibliography{references}

\end{document}